\documentclass{article}
\newtheorem{Theorem}{Theorem}

\newtheorem{Corollary}[Theorem]{Corollary}
\newtheorem{Definition}[Theorem]{Definition}

\begin{document}
\title{Secure Message Transmission in the Presence of a Fully Generalised Adversary}
\author{Chris Dowden \\
London School of Economics}
\maketitle
\setlength{\unitlength}{1cm}

\begin{abstract}
We investigate the problem of secure message transmission
in the presence of a `fully generalised' adversary,
who disrupts and listens to separate sets of communication wires.
We extend previous results by considering the case when these sets may have arbitrary size,
providing necessary and sufficient conditions for both one-way and two-way communication.
\end{abstract}

\section{Introduction} \label{intro}

A sender, $S$, and a receiver, $R$, are connected by several parallel communication wires
(these might, for example, represent vertex-disjoint paths from $S$ to $R$ in some underlying network).
We shall sometimes assume that these wires only permit one-way communication from $S$ to $R$;
other times, we shall assume that the wires allow two-way communication.

$S$ wishes to transmit a secret numerical message, $m$, to $R$.
Unfortunately, an adversary has control of some unknown subset of the wires,
meaning that he can listen to anything transmitted along particular wires
and can also disrupt transmissions so that something different is received.

Our challenge is to design a protocol for $S$ and $R$ to follow that is `secure',
in the sense that it is both (a) `reliable' ($R$ receives $m$ correctly)
and (b) `private' (the adversary must not be allowed to learn anything at all about $m$).

Results for the standard model where the adversary completely controls an unknown set of $k$ wires (for known $k$)
can be found in \cite{dol},
both for the case of one-way and two-way communication wires.
In \cite{kum} and \cite{des},
this work is then generalised to a model where the adversary chooses between sets of wires of different sizes.

Another extension that is also discussed in \cite{dol}
concerns a scenario where the wires listened to may differ from those that can be disrupted.
This work has connections to problems involving multiple adversaries,
where the overall set of disrupted wires may differ from the set
that can be listened to by any one party.

Previous results for this latter model all focus on the case where the set of wires listened to
and the set of wires disrupted both have given sizes.
In this paper,
we now extend this to a `fully generalised' model akin to the work of \cite{kum} and \cite{des},
where the adversary chooses between (pairs of) sets of different sizes.

The paper is structured as follows:
in Section~\ref{summary},
we summarise the previous relevant work;
in Section~\ref{1waysmt},
we look at the one-way case;
and in Sections~\ref{2waysmtcomp} and~\ref{2waysmtnotcomp},
we investigate two particular variants of the two-way case.

\section{Summary of Previous Results} \label{summary}

In this section,
we shall summarise relevant known results.
We start (in Subsection~\ref{defns}) with some basic definitions,
before then looking at the cases of a `$k$-adversary' (Subsection~\ref{k-ad}),
a `general adversary structure' (Subsection~\ref{genad})
and a `fully generalised adversary' (Subsection~\ref{fullygen}).
In the rest of this paper,
we shall then present new results for this latter case.

\subsection{Basic Definitions} \label{defns}

In this subsection,
we provide some basic definitions concerning the nature of the adversary and the required communications protocol.

We start with details of the influence of the adversary:

\begin{Definition}

We say that an adversary \textbf{disrupts} a communication wire to mean that he replaces the transmission on this wire with a new message,
so that the recipient only receives the new version.

We say that an adversary \textbf{listens} to a communication wire 
to mean that he hears everything transmitted (by any party) on this wire.

We say that an adversary \textbf{controls} a communication wire
to mean that he can both disrupt and listen to this wire.

\end{Definition}

Note that we do not assume that an adversary must disrupt all wires under his control ---
he may decide to allow some transmissions to pass through unaltered.
Also,
unless otherwise stated,
we do not distinguish between the case when the new transmission on a disrupted wire is specifically chosen by the adversary
and the case when the new transmission is random noise.

As mentioned in Section~\ref{intro},
the aim is for a sender $S$ to convey a secret numerical message $m$ to a receiver $R$
by using a pre-determined `secure' protocol:

\begin{Definition}

We say that a communications protocol is \textbf{reliable}
if $R$ can always deduce $m$ correctly.

We say that a communications protocol is \textbf{private}
if the adversary does not learn anything at all about $m$.

We say that a communications protocol is \textbf{secure}
if it is both reliable and private.

\end{Definition}

We assume throughout that the protocol is also known to the adversary.
For technical reasons,
we also implicitly assume that all arithmetic operations are carried out modulo some publically known large prime.

Before moving on, 
let us just mention in passing that there are also interesting probabilistic analogues to our definitions,
involved in the study of `almost secure' message transmission
(see, for example, \cite{cho}, \cite{fra}, \cite{saf} and \cite{sri}).
However, this is not to be the focus of the current paper.

\subsection{The $\mathbf{k}$-Adversary} \label{k-ad}

In this subsection,
we shall discuss the scenario of a `$k$-adversary',
providing details of results for secure message transmission
in both the case when the communication wires only permit one-way communication from $S$ to $R$
and the case when the wires allow unlimited two-way communication back and forth between $S$ and $R$.

We start with a definition:

\begin{Definition}

A \textbf{$\mathbf{k}$-adversary} can control any set of $k$ communication wires.
This is sometimes also known as a \textbf{threshold adversary with parameter $\mathbf{k}$}.

\end{Definition}

Throughout this paper,
we shall assume that the adversary's choice of wires is made before the start of the communications process
and then remains fixed.
We also assume that the identity of the chosen wires is unknown to $S$ and $R$.

The following result provides a complete description of the possibility of secure message transmission
using one-way communication wires:

\begin{Theorem} \textbf{(see \cite{dol})} \label{k1way}
In the presence of a $k$-adversary,
one-way secure message transmission using $n$ communication wires
is possible if and only if $n>3k$.
\end{Theorem}

Note that one successful protocol is for $S$ to choose a random degree $k$ polynomial $p(x)$ satisfying $p(0)=m$,
and transmit $p(i)$ on wire $i$ for all $i \in \{1,2, \ldots, 3k+1\}$.
We shall use this idea later.

There is also an analogous result concerning two-way communication wires.
Although we now allow unlimited communication through the wires,
so that information can be sent back and forth many times before $R$ eventually obtains $m$,
it turns out that it suffices just to have two communication rounds
(one from $R$ to $S$ and then one from $S$ to $R$).

\begin{Theorem} \textbf{(see \cite{dol})} \label{k2way}
In the presence of a $k$-adversary,
two-way secure message transmission using $n$ communication wires
is possible if and only if $n>2k$.
Furthermore,
when this condition is met,
secure message transmission in just two communication rounds is actually possible.
\end{Theorem}

\subsection{The General Adversary Structure} \label{genad}

In this subsection,
we shall extend the results of Subsection~\ref{k-ad}
to the case of a `general adversary structure',
where the possibilities for the set of wires controlled by the adversary may differ in size,
rather than all having cardinality $k$.
We shall again see results for both the one-way and two-way cases,
but first we start with an appropriate definition:

\begin{Definition}

Let $\mathcal{A}$ be a collection of sets of communication wires.
Then we say that an adversary has \textbf{general adversary structure} $\mathcal{A}$
to mean that the adversary can control any set of communication wires in $\mathcal{A}$.

\end{Definition}

For example,
if $\mathcal{A} = \{ \{1,2,3\}, \{1,2,4\}, \{1,5\} \}$
then the adversary would have a choice between controlling $\{1,2,3\}$ or $\{1,2,4\}$ or $\{1,5\}$.
The $k$-adversary is consequently the special case where $\mathcal{A}$ consists of precisely the sets of size $k$.

The following result deals with the one-way scenario:

\begin{Theorem} \textbf{(see~\cite{des})} \label{gen1way}
Let $\mathcal{A}$ be a general adversary structure.
Then one-way secure message transmission is possible if and only if no three sets in $\mathcal{A}$ cover all communication wires.
\end{Theorem}

There is also a two-way version:

\begin{Theorem} \textbf{(see \cite{kum})} \label{gen2way}
Let $\mathcal{A}$ be a general adversary structure.
Then two-way secure message transmission is possible if and only if no two sets in $\mathcal{A}$ cover all communication wires.
\end{Theorem}

The proof given in \cite{kum} uses induction,
and the successful protocol takes $|\mathcal{A}| - 1$ communication rounds between $S$ and $R$,
unlike the two-round result of Theorem~\ref{k2way}.
In our new work in Sections~\ref{1waysmt} and~\ref{2waysmtcomp},
we shall use a modified version of this induction procedure,
and in Corollary~\ref{2roundcor}
we shall actually obtain a successful two-round protocol for Theorem~\ref{gen2way}.

\subsection{The Fully Generalised Adversary} \label{fullygen}

We shall now further extend the definitions of Subsection~\ref{genad}
to deal with the scenario of a `fully generalised' adversary,
where the wires disrupted may differ from the wires listened to.
We shall also provide known one-way and two-way results for a special case of this type of adversary.
In the remainder of the paper,
we shall then extend these results to cover any fully generalised adversary.

We start with a definition:

\begin{Definition} 

Given two sets of communication wires $D$ and $L$
(which may or may not intersect),
we say that an adversary \textbf{controls} $(D,L)$ 
to mean that he can both disrupt all wires in $D$ and simultaneously listen to all wires in $L$.

Given a collection $\mathcal{A} = \{ (D_{1},L_{1}), (D_{2},L_{2}), \ldots, (D_{|\mathcal{A}|}, L_{|\mathcal{A}|}) \}$
of pairs of sets of communication wires,
we say that an adversary is a \textbf{fully generalised adversary}
with \textbf{fully generalised adversary structure} $\mathcal{A}$
to mean that the adversary can control any pair $(D_{i}, L_{i})$ in $\mathcal{A}$.

\end{Definition}

For example,
if $\mathcal{A} = \big\{ \big(\{1,2\}, \{1,4\}\big), \big(\{1,2,3\}, \{1\}\big) \big\}$
then the adversary would have a choice between being able to disrupt $\{1,2\}$ and also listen to $\{1,4\}$
or being able to disrupt $\{1,2,3\}$ and also listen to $\{1\}$.
Note that the general adversary structure discussed in Subsection~\ref{genad} 
is a special case of the fully generalised structure
where each pair $(D_{i}, L_{i})$ consists of two identical sets.
Consequently,
the $k$-adversary is also a special case of the fully generalised adversary.

Throughout this paper,
we shall always assume that our adversary is `oblivious',
in the following sense:

\begin{Definition} \label{obdef}

We say that a fully generalised adversary is \textbf{oblivious}
if he cannot hear the original message on a disrupted wire
unless he happens to also be listening to that wire.

\end{Definition}

Note that it was not necessary to distinguish between oblivious and non-oblivious adversaries in Subsections~\ref{k-ad} and~\ref{genad},
since the adversaries discussed there would always be listening to any disrupted wires.
Observe also that the non-oblivious case for a fully generalised adversary structure
$\mathcal{A} = \{ (D_{1},L_{1}), (D_{2},L_{2}), \ldots, (D_{|\mathcal{A}|}, L_{|\mathcal{A}|}) \}$
is essentially the same as the oblivious case for a fully generalised adversary structure
$\mathcal{A}^{\prime} = 
\{ (D_{1},L_{1} \cup D_{1}), (D_{2},L_{2} \cup D_{2}), \ldots, (D_{|\mathcal{A}|}, L_{|\mathcal{A}|} \cup D_{|\mathcal{A}|}) \}$,
and so non-oblivious results can be deduced by studying the oblivious case.

In addition to Definition~\ref{obdef},
we shall also now find it necessary to make the following definition:

\begin{Definition}

We say that an oblivious fully generalised adversary is \textbf{completely oblivious}
if he also cannot hear the new message on a disrupted wire
unless he happens to be listening to that wire.

\end{Definition}

Note that the completely oblivious case could correspond to random noise that is not heard by the adversary,
while the non-completely oblivious case could correspond either to 
a scenario where the new transmission on a disrupted wire is specifically chosen by the adversary
or to a scenario of random noise that is heard by the adversary.

As mentioned,
we shall now provide two known results for the fully generalised adversary,
both for the special case when $\mathcal{A}$ consists of precisely those pairs $(D,L)$ for which $|D|=d$ and $|L|=l$.
We start with the one-way case:

\begin{Theorem} \textbf{(see \cite{dol})} \label{dl1way}
Given fixed constants $d$ and $l$,
let $\mathcal{A}$ be the fully generalised adversary structure consisting of precisely those pairs $(D,L)$ for which $|D|=d$ and $|L|=l$,
and suppose that the adversary is oblivious.
Then one-way secure message transmission using $n$ communication wires is possible 
if and only if $n > 2d+l$.
\end{Theorem}

Note that since the communication in Theorem~\ref{dl1way} is one-way,
it does not matter whether or not the adversary is completely oblivious.

Observe also that Theorem~\ref{dl1way} is a direct extension of Theorem~\ref{k1way},
but not Theorem~\ref{gen1way}.
In Section~\ref{1waysmt},
we shall present a new result that deals with any fully generalised adversary,
thus incorporating both Theorem~\ref{gen1way} and Theorem~\ref{dl1way}.

We first conclude this section with an analogous result to Theorem~\ref{dl1way} for two-way communication:

\begin{Theorem} \textbf{(see \cite{dol})} \label{dl2way}
Given fixed constants $d$ and $l$,
let $\mathcal{A}$ be the fully generalised adversary structure consisting of precisely those pairs $(D,L)$ for which $|D|=d$ and $|L|=l$,
and suppose that the adversary is completely oblivious.
Then two-way secure message transmission using $n$ communication wires is possible 
if and only if $n > d + \max\{d,l\}$.
Furthermore,
when this condition is met,
secure message transmission in just two communication rounds is actually possible.
\end{Theorem}

Note that this time we assume that the adversary is completely oblivious.
In Section~\ref{2waysmtnotcomp},
we shall present new work for the non-completely oblivious case,
obtaining a different result.

Observe also that Theorem~\ref{dl2way} is a direct improvement on Theorem~\ref{k2way},
but not Theorem~\ref{gen2way}.
In Section~\ref{2waysmtcomp},
we shall present a new result dealing with any completely oblivious fully generalised adversary structure,
incorporating both Theorem~\ref{gen2way} and Theorem~\ref{dl2way}.

\section{One-Way Secure Message Transmission} \label{1waysmt}

In Section~\ref{intro},
we gave details of previous relevant work on secure message transmission.
In the remainder of this paper,
we shall now present new results,
starting here with the one-way case.

Recall that we have already seen a necessary and sufficient condition
for successful one-way communication in the presence of a $k$-adversary (Theorem~\ref{k1way}),
as well as two separate extensions of this involving 
a general adversary structure (Theorem~\ref{gen1way})
and a special case of the fully generalised adversary (Theorem~\ref{dl1way}).
In Theorem~\ref{main1way},
we shall now unify these latter two results to deal with any fully generalised adversary:

\begin{Theorem} \label{main1way}
Let $\mathcal{A} = \{ (D_{1},L_{1}), (D_{2},L_{2}), \ldots, (D_{|\mathcal{A}|}, L_{|\mathcal{A}|}) \}$
be a fully generalised adversary structure,
and suppose that the adversary is oblivious.
\footnote{As with Theorem~\ref{dl1way},
it does not matter whether or not the adversary is completely oblivious.}
Then one-way secure message transmission is possible if and only if
there exist no $D_{i}$, $D_{j}$ and $L_{k}$
(including the possibility that $k$ equals $i$ or $j$)
such that $D_{i} \cup D_{j} \cup L_{k}$ covers all communication wires.
\end{Theorem}
\textbf{Proof}
Let us begin by assuming that there exist no $D_{i}$, $D_{j}$ and $L_{k}$
such that $D_{i} \cup D_{j} \cup L_{k}$ covers all communication wires,
and let us provide a protocol for one-way secure message transmission.

We shall proceed by using induction on $|\mathcal{A}|$.
The inductive step will require $|\mathcal{A}| \geq 4$,
and so the base case will be $|\mathcal{A}|=3$,
i.e.~$\mathcal{A} = \{ (D_{1},L_{1}), (D_{2},L_{2}), (D_{3},L_{3}) \}$
(note that this will then also deal with the case $|\mathcal{A}|<3$,
since we could take some of the $(D_{i},L_{i})$ to be $(\emptyset, \emptyset)$).

\textit{Base case ($|\mathcal{A}|=3)$:}
By the conditions of the theorem,
there exist (not necessarily distinct) wires
$w_{1} \notin (D_{1} \cup D_{2} \cup L_{1})$, $w_{2} \notin (D_{1} \cup D_{2} \cup L_{2})$ and  $w_{3} \notin (D_{1} \cup D_{2} \cup L_{3})$.
Let $S$ choose two random numbers $r_{1}$ and $r_{2}$,
and send $r_{1}$ on wire $w_{1}$, $r_{2}$ on wire $w_{2}$ and $m-r_{1}-r_{2}$ on wire $w_{3}$.
Observe that $m$ will still be a secret to the adversary,
since he will only be able to listen to at most two of the transmissions,
and that $R$ will be able to compute $m$ correctly unless the adversary  chooses $(D_{3},L_{3})$.
Let $S$ then also use the same procedure (with different random numbers) with wires not in $D_{1} \cup D_{3}$,
and then with wires not in $D_{2} \cup D_{3}$.
Each time,
$m$ will remain private,
and $R$ will be able to compute it correctly unless the adversary chooses $(D_{2},L_{2})$ or, respectively, $(D_{1},L_{1})$.
Hence, in total,
$R$ will compute three values,
at least two of which will be the same --- this value must be $m$.

\textit{Inductive step ($|\mathcal{A}|=l>3)$:}
Suppose now that a successful protocol has been established whenever $|\mathcal{A}|<l$,
and let us deal with the case when $|\mathcal{A}|=l$.
Let us create four new adversary structures $\mathcal{A}_{1}$, $\mathcal{A}_{2}$, $\mathcal{A}_{3}$ and $\mathcal{A}_{4}$
such that $\mathcal{A}_{j} = \mathcal{A} \setminus (D_{j},L_{j})$
(i.e.~$\mathcal{A}_{1} = \{ (D_{2}, L_{2}), (D_{3}, L_{3}), \ldots, (D_{l}, L_{l}) \}$, etc).
Since $|\mathcal{A}_{j}| = l-1$ for all $j$,
we know by the induction hypothesis that for each $\mathcal{A}_{j}$ there exists a successful protocol $\psi_{j}$.
It is crucial to note that if $\psi_{j}$ is used in the presence of the complete adversary structure $\mathcal{A}$,
then transmission will be secure unless the adversary chooses $(D_{j}, L_{j})$. 
Hence, if $S$ chooses a random number $r$ and,
for each $j \in \{ 1,2,3,4 \}$,
uses $\psi_{j}$ to transmit $m+jr$,
then all but at most one of these four transmissions will be secure.
Thus, as with the successful four wire protocol in the presence of a $1$-adversary
(see the comment following Theorem~\ref{k1way}),
$R$ can then deduce $m$ without the adversary gaining any knowledge himself. \\

Let us now look at the converse direction.
Suppose that there exist $D_{i}$, $D_{j}$ and $L_{k}$
such that $D_{i} \cup D_{j} \cup L_{k}$ covers all communication wires,
and, with the aim of achieving a contradiction,
let us also suppose that there exists a successful protocol, $\psi$,
for one-way secure message transmission.

Note that the information transmitted by $S$ through the wires depends not only on the message and the protocol,
but also perhaps on various random choices that $S$ has to take as part of the protocol
(e.g.~in the third paragraph of this proof,
$S$ had to choose random numbers $r_{1}$ and $r_{2}$).
Let us suppose that $S$ wishes to send a message $m_{1}$ using protocol $\psi$,
and that he also makes random choices $\mathcal{C}_{1}$ as part of the protocol.

Consider the information transmitted through the wires in $L_{k}$.
By privacy,
this information must be compatible with all possible messages,
i.e.~given any possible message $m_{2}$ there must exist random choices $\mathcal{C}_{2}$
such that sending message $m_{2}$ using protocol $\psi$ with random choices $\mathcal{C}_{2}$
results in exactly the same transmissions in $L_{k}$ as sending message $m_{1}$ using protocol $\psi$ with random choices $\mathcal{C}_{1}$.

One possibility for the adversary (let us call this ``Possibility A") is to choose $(D_{i}, L_{i})$
and disrupt $D_{i}$ in such a way that the information received by $R$ through the wires in $D_{i}$
happens to be exactly what would be transmitted by $S$
if $S$ was sending $m_{1}$ using protocol $\psi$ with random choices $\mathcal{C}_{1}$.

Another possiblity for the adversary (let us call this ``Possibility B") is to instead choose $(D_{j}, L_{j})$
and disrupt $D_{j} \setminus D_{i}$ 
in such a way that the information received by $R$ through the wires in $D_{j} \setminus D_{i}$ happens to be exactly what would be transmitted by $S$
if $S$ was sending $m_{2}$ using protocol $\psi$ with random choices $\mathcal{C}_{2}$.

Note that if Possibility A occurs and S sends $m_{2}$ using protocol $\psi$ with random choices $\mathcal{C}_{2}$,
then $R$ will receive exactly the same transmissions as if Possibility B occurs
and $S$ sends $m_{1}$ using protocol $\psi$ with random choices $\mathcal{C}_{1}$.
Hence, it will be impossible for $R$ to determine the message reliably,
and so we have obtained our desired contradiction.
\phantom{qwerty}
\setlength{\unitlength}{.25cm}
\begin{picture}(1,1)
\put(0,0){\line(1,0){1}}
\put(0,0){\line(0,1){1}}
\put(1,1){\line(-1,0){1}}
\put(1,1){\line(0,-1){1}}
\end{picture} \\

\section{Two-Way Secure Message Transmission with a Completely Oblivious Adversary} \label{2waysmtcomp}

In Section~\ref{1waysmt},
we presented a necessary and sufficient condition for one-way secure message transmission
in the presence of an oblivious adversary,
and noted that it did not matter whether or not the adversary is completely oblivious.
We shall now proceed with the two-way case,
for which we shall find that we do now need to make such a distinction.
Consequently,
this section will focus on the completely oblivious scenario,
while Section~\ref{2waysmtnotcomp} will concern the non-completely oblivious analogue.

Recall that we have already seen a necessary and sufficient condition with the $k$-adversary (Theorem~\ref{k2way}),
as well as extensions of this to the general adversary (Theorem~\ref{gen2way})
and a special case of the fully generalised adversary (Theorem~\ref{dl2way}).
In Theorem~\ref{maincompobliv},
we shall now combine these latter two results to include any fully generalised adversary.

Furthermore,
the successful protocol will only take two communication rounds,
allowing us to obtain (in Corollary~\ref{2roundcor})
an improvement to the protocol given in the proof of Theorem~\ref{gen2way}.

\begin{Theorem} \label{maincompobliv}
Let $\mathcal{A} = \{ (D_{1},L_{1}), (D_{2},L_{2}), \ldots, (D_{|\mathcal{A}|}, L_{|\mathcal{A}|}) \}$
be a fully generalised adversary structure,
and suppose that the adversary is completely oblivious.
Then two-way secure message transmission is possible if and only if
there exist no $D_{i}$ and $D_{j}$
such that $D_{i} \cup D_{j}$ covers all communication wires
\emph{and} no $D_{i}$ and $L_{j}$
(including the possibility that $i$ equals $j$)
such that $D_{i} \cup L_{j}$ covers all communication wires.
Furthermore, when these conditions are met,
secure message transmission in just two communication rounds is actually possible.
\end{Theorem}
\textbf{Proof}
Let us start by using induction to provide a successful two-round protocol for when the adversary structure does satisfy the conditions of the theorem.
As in the proof of Theorem~\ref{main1way},
the inductive step will require $|\mathcal{A}| \geq 4$
and so the base case will be $|\mathcal{A}|=3$
(recall that this automatically incorporates the case $|\mathcal{A}| < 3$). \\

\textit{Base case ($|\mathcal{A}|=3)$:} 

By the conditions of the theorem,
there exist (not necessarily distinct) wires 
$w_{1,1} \notin (D_{1} \cup L_{1})$,
$w_{1,2} \notin (D_{1} \cup L_{2})$,
$w_{1,3} \notin (D_{1} \cup L_{3})$,
$w_{2,1} \notin (D_{2} \cup L_{1})$,
$w_{2,2} \notin (D_{2} \cup L_{2})$,
$w_{2,3} \notin (D_{2} \cup L_{3})$,
$w_{3,1} \notin (D_{3} \cup L_{1})$,
$w_{3,2} \notin (D_{3} \cup L_{2})$,
$w_{3,3} \notin (D_{3} \cup L_{3})$,
$w_{4,1} \notin (D_{1} \cup D_{2})$,
$w_{4,2} \notin (D_{1} \cup D_{3})$
and $w_{4,3} \notin (D_{2} \cup D_{3})$.

\textit{Step $1$:}
Let $R$ choose four random degree-one polynomials $p_{1}(x)$, $p_{2}(x)$, $p_{3}(x)$ and $p_{4}(x)$.
For each polynomial $p_{i}(x)$,
let $R$ then also choose six random numbers $r_{i,j,k}$,
where $j \in \{ 1,2,3 \}$ and $k \in \{ 1,2 \}$
(so there will be twenty-four random integers in total).
Then, for each polynomial $p_{i}(x)$ and all $j \in \{ 1,2,3 \}$,
let $R$ send $r_{i,j,1}$ on wire $w_{j,1}$,
$r_{i,j,2}$ on wire $w_{j,2}$,
$p_{i}(j) - r_{i,j,1} - r_{i,j,2}$ on wire $w_{j,3}$,
and $p_{i}(4)$ on all three wires $w_{4,1}$, $w_{4,2}$ and $w_{4,3}$.

Note that, for each polynomial $p_{i}(x)$ and all $j \in \{ 1,2,3 \}$,
$p_{i}(j)$ will be equal to the sum of the transmissions on wires $w_{j,1}$, $w_{j,2}$ and $w_{j,3}$.
Hence, (i) $p_{i}(1)$ cannot be eavesdropped and can only be disrupted if the adversary chooses $(D_{2},L_{2})$ or $(D_{3},L_{3})$,
(ii) $p_{i}(2)$ cannot be eavesdropped and can only be disrupted if the adversary chooses $(D_{1},L_{1})$ or $(D_{3},L_{3})$,
and (iii) $p_{i}(3)$ cannot be eavesdropped and can only be disrupted if the adversary chooses $(D_{1},L_{1})$ or $(D_{2},L_{2})$.
Additionally, observe that (iv) $p_{i}(4)$ cannot be disrupted (since it can be deduced by majority vote), but can be eavesdropped.

Let $\overline{p_{i}(j)}$ denote the value of $p_{i}(j)$ received by $S$ (so $\overline{p_{i}(4)} = p_{i}(4)$).

\textit{Step $2$, case (A):}
If there exists an $i$ such that $\{ \overline{p_{i}(1)}, \overline{p_{i}(2)}, \overline{p_{i}(3)}, p_{i}(4) \}$ is consistent with a degree-one polynomial,
then note that this polynomial must be $p_{i}(x)$ and must still be unknown to the adversary,
so $S$ can simply transmit $i$ and $m+p_{i}(0)$ `publically'
(i.e.~on wires $w_{4,1}$, $w_{4,2}$ and $w_{4,3}$, with $R$ then using majority vote)
and $R$ can deduce $m$.
\footnote{Clearly, $p_{i}(x)$ will still be unknown to the adversary,
but note that this would not necessarily be the case 
if he were not completely oblivious
and the new values on disrupted wires were known to him.
For example, it would be possible to have happened to have disrupted $p_{i}(1)$, say,
with the correct value and to have also eavesdropped $p_{i}(4)$,
in which case the very fact that $S$ is transmitting $m+p_{i}(0)$ at all
(rather than following the procedures of cases (B) or (C))
would then allow the adversary to deduce what had occurred!}

\textit{Step $2$, case (B) (to be used if the conditions of case (A) do not hold):}
If there exists an $i$ such that no \textit{three} values of $\{ \overline{p_{i}(1)}, \overline{p_{i}(2)}, \overline{p_{i}(3)}, p_{i}(4) \}$ are consistent with the same degree-one polynomial,
then this means that two of the values in $\{ p_{i}(1), p_{i}(2), p_{i}(3) \}$ have been disrupted.
Hence, $S$ can transmit $i$, $\overline{p_{i}(1)}$,  $\overline{p_{i}(2)}$ and $\overline{p_{i}(3)}$ `publically',
allowing $R$ to deduce \textit{exactly} which set the adversary chose.
$S$ can then select another polynomial $p_{j}(x)$,
compute the three possibilities $p^{\prime}_{j}(0)$, $p^{\prime \prime}_{j}(0)$ and $p^{\prime \prime \prime}_{j}(0)$ for $p_{j}(0)$
(depending on whether $\overline{p_{j}(1)} = p_{j}(1)$, $\overline{p_{j}(2)} = p_{j}(2)$ or $\overline{p_{j}(3)} = p_{j}(3)$, respectively)
and `publically' send $j$, $m+p^{\prime}_{j}(0)$, $m+p^{\prime \prime}_{j}(0)$ and $m+p^{\prime \prime \prime}_{j}(0)$,
allowing $R$ to deduce $m$.

\textit{Step $2$, case (C) (to be used if the conditions of cases (A) and (B) do not hold):}
Suppose that for all $i$,
there exist exactly three values of $\{ \overline{p_{i}(1)}, \overline{p_{i}(2)}, \overline{p_{i}(3)}, p_{i}(4) \}$ that are consistent with the same degree-one polynomial.
Observe that these three values must include $p_{i}(4)$,
and so the value not included must be either $\overline{p_{i}(1)}$, $\overline{p_{i}(2)}$ or $\overline{p_{i}(3)}$,
i.e.~three alternatives.
Since there are four values of $i$,
the pigeon-hole principle then implies that there must exist two values of $i$,
let us call these $i_{1}$ and $i_{2}$,
that share the `same' alternative ---
without loss of generality let us suppose that this is the first alternative,
i.e.~$\overline{p_{i_{1}}(1)}$ and $\overline{p_{i_{2}}(1)}$.

Then we know that in polynomial $p_{i_{1}}(x)$,
either (a) $p_{i_{1}}(1)$ was disrupted
or (b) $p_{i_{1}}(2)$ and $p_{i_{1}}(3)$ were both disrupted.
Similarly, in polynomial $p_{i_{2}}(x)$,
either (a) $p_{i_{2}}(1)$ was disrupted
or (b) $p_{i_{2}}(2)$ and $p_{i_{2}}(3)$ were both disrupted.
Crucially, note (from the second paragraph of Step $1$)
that (a) is true for $p_{i_{1}}(x)$ if and only if (a) is also true for $p_{i_{2}}(x)$.

Hence, if $S$ transmits $i_{1}$, $\overline{p_{i_{1}}(1)}$, $\overline{p_{i_{1}}(2)}$ and $\overline{p_{i_{1}}(3)}$ `publically',
this actually allows $R$ to deduce whether (a) or (b) is true for $p_{i_{2}}(x)$.
$S$ can then compute the two possibilities $p^{\prime}_{i_{2}}(0)$ and $p^{\prime \prime}_{i_{2}}(0)$  for $p_{i_{2}}(0)$
(depending on whether (a) is true or (b) is true)
and also `publically' send $i_{2}$, $m+p^{\prime}_{i_{2}}(0)$ and $m+p^{\prime \prime}_{i_{2}}(0)$,
enabling $R$ to then deduce $m$. \\

\textit{Inductive step ($|\mathcal{A}|=l \geq 4)$:} 

Suppose now that a successful protocol has been established whenever $|\mathcal{A}| <l$,
and let us deal with the case when $|\mathcal{A}|=l$.

As in the proof of Theorem~\ref{main1way},
let us consider the following four adversary structures of size $l-1$:
$\mathcal{A}_{1} = \mathcal{A} \setminus (D_{1},L_{1})$,
$\mathcal{A}_{2} = \mathcal{A} \setminus (D_{2},L_{2})$,
$\mathcal{A}_{3} = \mathcal{A} \setminus (D_{3},L_{3})$,
and $\mathcal{A}_{4} = \mathcal{A} \setminus (D_{4},L_{4})$.
By the induction hypothesis,
for each $A_{j}$ there exists a successful two-round protocol $\psi_{j}$
(with one round from $R$ to $S$ and then one round from $S$ to $R$).
As in the proof of Theorem~\ref{main1way},
note that if $\psi_{j}$ is used in the presence of the complete adversary structure $\mathcal{A}$,
then transmission will be secure unless the adversary chooses $(D_{j}, L_{j})$.
Hence, if protocol $\psi_{j}$ is used to transmit $m+jr$ for all $j \in \{ 1,2,3,4 \}$
(where $r$ is some random integer chosen by $S$ and $m$ is the secret message),
then all but at most one of these transmissions will be secure.
Thus, as in the proof of Theorem~\ref{main1way},
$R$ can then deduce $m$ without the adversary gaining any knowledge himself. \\

Now we shall show that secure message transmission is not possible at all 
(in any number of rounds)
if either of the conditions of the theorem are not met.

First, let us deal with the case when there exist $D_{i}$ and $D_{j}$ such that $D_{i} \cup D_{j}$ covers all communication wires.
The proof follows the same argument as that given for Theorem 5.2 in \cite{dol} (our Theorem~\ref{dl2way}).

Suppose, aiming for a contradiction,
that there exists a successful protocol,
and let $m_{1}$ and $m_{2}$ be two potential messages.
Now consider the first round of transmissions sent by $S$ ---
suppose that $\alpha_{k}$ is transmitted on wire $k$ if $m_{1}$ is the message
and $S$ makes random choices $\mathcal{C}_{1}$
and that $\beta_{k}$ is transmitted on wire $k$ if $m_{2}$ is the message
and $S$ makes random choices $\mathcal{C}_{2}$.

One possible strategy for the adversary
(let us call this ``Possibility A")
is to choose $(D_{i},L_{i})$ and disrupt all wires in $D_{i}$,
with the possibility that the new transmission on wire $k$ (for all $k \in D_{i}$) is now $\beta_{k}$.
Similarly, another possible strategy for the adversary (``Possibility B'')
is to choose $(D_{j},L_{j})$ and disrupt all wires in $D_{j} \setminus D_{i}$,
with the possibility that the new transmission on wire $k$ (for all $k \in D_{j} \setminus D_{i}$) is now $\alpha_{k}$.

Note that if the message is $m_{1}$,
$S$ makes random choices $\mathcal{C}_{1}$, and Possibility $A$ occurs,
then $R$ will receive exactly the same transmissions as if
the message is $m_{2}$,
$S$ makes random choices $\mathcal{C}_{2}$, and Possibility $B$ occurs.
Hence, the next round of transmissions sent by $R$ would have to be the same in both cases.

By applying this same argument to every pair of transmission rounds,
we see that it is possible for the adversary to force the protocol to run forever without $R$ ever being able to deduce the message,
and so we achieve our desired contradiction.

Now let us deal with the case when there exist $D_{i}$ and $L_{j}$
(including the possibility that $i$ equals $j$)
such that $D_{i} \cup L_{j}$ covers all communication wires.
Again, the proof follows the same argument as that given for Theorem 5.2 in \cite{dol}.

Suppose, aiming for a contradiction,
that there exists a successful protocol,
and let $m$ and $m^{\prime}$ be potential messages.
Let $\alpha_{k}^{l}$ denote the transmission on wire $k$ in round $l$ when $m$ is the message,
random choices $\mathcal{C}_{S}$ and $\mathcal{C}_{R}$ are made by $S$ and $R$, respectively,
and no wires are ever disrupted.
By privacy,
there must exist random choices $\mathcal{C}_{S}^{\prime}$ and $\mathcal{C}_{R}^{\prime}$ such that
if $m^{\prime}$ is the message, choices $\mathcal{C}_{S}^{\prime}$ and $\mathcal{C}_{R}^{\prime}$ are made,
and no wires are ever disrupted,
then exactly the same transmissions in $L_{j}$ are made.
Hence, if we let $\beta_{k}^{l}$ denote the transmissions on wire $k$ in round $l$ when $m^{\prime}$ is the message
(given choices $\mathcal{C}_{S}^{\prime}$ and $\mathcal{C}_{R}^{\prime}$ and no disruption of wires),
then we have $\beta_{k}^{l} = \alpha_{k}^{l}$ for all $l$ for all $k \in L_{j}$.

Without loss of generality,
let us suppose that the first round of transmissions is from $S$ to $R$,
and let us consider the scenario when $m$ is the message,
$S$ makes random choices $C_{S}$,
and $R$ makes random choices $C_{R}^{\prime}$.

A strategy for the adversary would be to disrupt all wires in $D_{i}$,
with the possibility that all transmissions over these wires could be altered to $\beta_{k}^{1}$.
Hence, since $R$ also receives $\beta_{k}^{1}$ ($=\alpha_{k}^{1}$) over all wires in $L_{j}$
(and $D_{i} \cup L_{j}$ covers every wire),
$R$ would then respond by transmitting $\beta_{k}^{2}$ on wire $k$ for all $k$ in round two.
One option now for the adversary would be to again disrupt all wires in $D_{i}$,
with the possibility that all $R$'s transmissions could be altered to $\alpha_{k}^{2}$.
Note that $S$ would then respond by transmitting $\alpha_{k}^{3}$ on wire $k$ for all $k$ in round three.
Continuing in this manner,
we can observe that $R$ would ultimately think that message $m^{\prime}$ had been sent, which would be wrong.
\phantom{qwerty}
\setlength{\unitlength}{.25cm}
\begin{picture}(1,1)
\put(0,0){\line(1,0){1}}
\put(0,0){\line(0,1){1}}
\put(1,1){\line(-1,0){1}}
\put(1,1){\line(0,-1){1}}
\end{picture} \\

As noted,
Theorem~\ref{maincompobliv} provides a successful protocol in just two communication rounds
when the conditions are met.
Consequently,
by taking $D_{i} = L_{i}$ for all $i$,
we also obtain the following two-round result 
for the case of a general adversary structure
(improving on the $|\mathcal{A}|-1$ round protocol given in the proof of Theorem~\ref{gen2way}).

\begin{Corollary} \label{2roundcor}
Let $\mathcal{A}$ be a general adversary structure,
and suppose that no two sets in $\mathcal{A}$ cover all communication wires.
Then secure message transmission in just two communication rounds is possible.
\end{Corollary}

\section{Two-Way Secure Message Transmission with a Non-Completely Oblivious Adversary} \label{2waysmtnotcomp}

In Section~\ref{2waysmtcomp},
we gave a necessary and sufficient condition for two-way secure message transmission
in the presence of a completely oblivious adversary,
and noted that only two communication rounds are required
whenever successful protocols are possible.
In this section,
we shall now look at the case of an adversary that is not completely oblivious,
providing (in Theorem~\ref{mainnotcompobliv})
a necessary and sufficient condition for the existence of two-round protocols,
which we shall see differs from the equivalent result for the completely oblivious case.

\begin{Theorem} \label{mainnotcompobliv}
Let $\mathcal{A} = \{ (D_{1},L_{1}), (D_{2},L_{2}), \ldots, (D_{|\mathcal{A}|}, L_{|\mathcal{A}|}) \}$
be a fully generalised adversary structure,
and suppose that the adversary is oblivious, but not completely oblivious.
Then two-way secure message transmission in just two communication rounds
is possible if and only if
there exist no $D_{i}$, $D_{j}$ and $L_{j}$
such that $D_{i} \cup D_{j} \cup L_{j}$ covers all communication wires.
\end{Theorem}
\textbf{Proof}
Let us start by showing that a successful two-round protocol exists
when the adversary structure satisfies the conditions of the theorem.

It will prove helpful to consider the adversary structure
$\mathcal{A}^{\prime} = 
\{ (D_{1},L_{1} \cup D_{1}), (D_{2},L_{2} \cup D_{2}), \ldots, (D_{|\mathcal{A}|}, L_{|\mathcal{A}|} \cup D_{|\mathcal{A}|}) \}$.
Note that any protocol that is successful in the presence of $\mathcal{A}^{\prime}$
must also be successful in the presence of $\mathcal{A}$,
since $\mathcal{A}^{\prime}$ is stronger.
Furthermore,
note that with $\mathcal{A}^{\prime}$
it is irrelevant whether the adversary is defined to be completely oblivious or not
(or even whether he is oblivious at all),
since all disrupted wires will also be listened to.

Hence,
by Theorem~\ref{maincompobliv},
a successful two-round protocol in the presence of $\mathcal{A}^{\prime}$
(and consequently $\mathcal{A}$)
is possible if there exist no $i$ and $j$ such that
$D_{i} \cup \{ L_{j} \cup D_{j} \}$ covers all communication wires.
\\

Let us now show that a successful two-round protocol
is not possible if the conditions of the theorem are not met.

The proof will be by contradiction,
so let us suppose that there does exist a successful two-round protocol.
To start with,
consider the case when no wires are disrupted.
Given a message $m$ and random choices $\mathcal{C}_{R}$ and $\mathcal{C}_{S}$ made by $R$ and $S$,
let us represent the transmissions sent in round $1$ (by $R$) by the vectors $\underline{\alpha_{1}}$, $\underline{\beta_{1}}$ and $\underline{\gamma_{1}}$,
where $\underline{\alpha_{1}}$, $\underline{\beta_{1}}$ and $\underline{\gamma_{1}}$ denote the transmissions sent on wires in $D_{i} \setminus L_{j}$, $D_{j} \setminus ( D_{i} \cup L_{j} )$ and $L_{j}$, respectively.
Similarly, let us represent the transmissions sent in round $2$ (by $S$) by $\underline{\alpha_{2}}$, $\underline{\beta_{2}}$ and $\underline{\gamma_{2}}$.

Now consider the possibility that the adversary could choose $(D_{j}, L_{j})$
and happen to disrupt all wires in $D_{j} \setminus ( D_{i} \cup L_{j} )$ with the correct values $\underline{\beta_{1}}$ and $\underline{\beta_{2}}$.
Privacy then implies that, for any message $m^{\prime}$, 
there must exist random choices $\mathcal{C}_{R}^{\prime}$ and $\mathcal{C}_{S}^{\prime}$ 
and vectors $\underline{\alpha_{1}}^{\prime}$, $\underline{\beta_{1}}^{\prime}$, $\underline{\alpha_{2}}^{\prime}$, $\underline{\beta_{2}}^{\prime}$
such that the transmissions sent in round $1$ (by $R$)
are $\underline{\alpha_{1}^{\prime}}$, $\underline{\beta_{1}^{\prime}}$ and $\underline{\gamma_{1}}$,
the transmissions received (by $S$) are  $\underline{\alpha_{1}^{\prime}}$, $\underline{\beta_{1}}$ and $\underline{\gamma_{1}}$,
the transmissions sent in round $2$ (by $S$) are $\underline{\alpha_{2}^{\prime}}$, $\underline{\beta_{2}^{\prime}}$ and $\underline{\gamma_{2}}$,
and the transmissions received (by $R$) are $\underline{\alpha_{2}^{\prime}}$, $\underline{\beta_{2}}$ and $\underline{\gamma_{2}}$.

One possible strategy for the adversary 
(let us call this ``Possibility A") is to choose $(D_{i}, L_{i})$
and happen to disrupt the wires in $D_{i} \setminus L_{j}$ with $\underline{\alpha_{1}^{\prime}}$ in round $1$ and $\underline{\alpha_{2}}$ in round $2$.
Another strategy (``Possibility B")  is to instead choose $(D_{j}, L_{j})$
and make no disruptions in round $1$,
but happen to disrupt the wires in $D_{j} \setminus ( D_{i} \cup L_{j} )$ with $\underline{\beta_{2}^{\prime}}$ in round $2$.

Now consider the possibility that $R$ makes choices $\mathcal{C}_{R}$ and transmits $\underline{\alpha_{1}}$, $\underline{\beta_{1}}$ and $\underline{\gamma_{1}}$ in round $1$.

Under Possibility A,
$S$ will receive $\underline{\alpha_{1}^{\prime}}$, $\underline{\beta_{1}}$ and $\underline{\gamma_{1}}$.
If $S$  wishes to send $m^{\prime}$ and makes random choices $\mathcal{C}_{S}^{\prime}$,
then he will transmit $\underline{\alpha_{2}^{\prime}}$, $\underline{\beta_{2}^{\prime}}$ and $\underline{\gamma_{2}}$,
and $R$ will receive $\underline{\alpha_{2}}$, $\underline{\beta_{2}^{\prime}}$ and $\underline{\gamma_{2}}$.

Under Possibility B,
$S$ will instead receive $\underline{\alpha_{1}}$, $\underline{\beta_{1}}$ and $\underline{\gamma_{1}}$.
If $S$ wishes to send $m$ and makes random choices $\mathcal{C}_{S}$,
then he will transmit $\underline{\alpha_{2}}$, $\underline{\beta_{2}}$ and $\underline{\gamma_{2}}$,
and $R$ will again receive $\underline{\alpha_{2}}$, $\underline{\beta_{2}^{\prime}}$ and $\underline{\gamma_{2}}$.

Hence, $R$ will be unable to distinguish between $m$ and $m^{\prime}$,
and so the protocol is not successful.
\phantom{qwerty}
\setlength{\unitlength}{.25cm}
\begin{picture}(1,1)
\put(0,0){\line(1,0){1}}
\put(0,0){\line(0,1){1}}
\put(1,1){\line(-1,0){1}}
\put(1,1){\line(0,-1){1}}
\end{picture} \\

\section{Concluding Remarks} \label{conc}

In this paper,
we have investigated secure message transmission
in the presence of a fully generalised adversary,
where the set of communication wires disrupted 
may differ from those listened to
and may be of any size.

We have extended previous results into a complete description of necessary and sufficient conditions
for both the one-way case (Theorem~\ref{main1way})
and the two-way case with a completely oblivious adversary (Theorem~\ref{maincompobliv}).

In the additional case of an oblivious adversary that is not completely oblivious,
we have presented a necessary and sufficient condition for the existence of a protocol that uses only two communication rounds
(Theorem~\ref{mainnotcompobliv}),
and seen that this differs from with a completely oblivious adversary.

The remaining issue for the additional case consequently concerns resolving the exact conditions
under which any successful protocol
(perhaps using more than two rounds)
is obtainable.
Here,
a sufficient condition is clearly given by the two-round result of Theorem~\ref{mainnotcompobliv},
while a necessary condition is implied by the completely oblivious case of Theorem~\ref{maincompobliv}.
It would be interesting to try to close this gap.

\section*{Acknowledgements}

This paper was written at Royal Holloway and Bedford New College (University of London)
as part of the EU-funded project `Internet of Energy for Electric Mobility',
under the supervision of Stephen Wolthusen,
and I am grateful for this support.
I would also like to thank the reviewers for taking the time to make detailed comments on my manuscript.


\begin{thebibliography}{99}

\bibitem{kum} M.~V.~N. Ashwin Kumar, P.~R. Goundan, K. Srinathan, C. Pandu Rangan,
\textit{On Perfectly Secure Communication over Arbitrary Networks},
Twenty-First ACM Symposium on Principles of Distributed Computing (2002),
193--202.

\bibitem{cho} A. Choudhury, K. Kurosawa, A. Patra,
\textit{Simple and Efficient Single Round Almost Perfectly Secure Message Transmission Tolerating Generalized Adversary},
Proceedings of the 9th International Conference on Applied Cryptography and Network Security (2011),
292--308.

\bibitem{des} Y. Desmedt, Y. Wang, M. Burmester,
\textit{A Complete Characterisation of Tolerable Adversary Structures for Secure Point-to-Point Transmissions Without Feedback},
Proceedings of the 16th International Symposium on Algorithms and Computation (2005),
277--287.

\bibitem{dol} D. Dolev, C. Dwork, O. Waarts, M. Yung,
\textit{Perfectly Secure Message Transmission},
Journal of the Association for Computing Machinery \textbf{40} (1993),
17--47.

\bibitem{fra} M. Franklin, R. Wright,
\textit{Secure Communication in Minimal Connectivity Models},
Journal of Cryptology \textbf{13} (2000),
9--30.

\bibitem{saf} R. Safavi-Naini, M.~A. Tuhin,
\textit{Bounds and Constructions for $1$-Round $(0,d)$-Secure Message Transmission against Generalized Adversary},
Proceedings of the 5th International Conference on Cryptology in Africa (2012),
82--98.

\bibitem{sri} K. Srinathan, A. Patra, A. Choudhary, C. Pandu Rangan,
\textit{Unconditionally Secure Message Transmission in Arbitrary Directed Synchronous Networks Tolerating Generalized Mixed Adversary},
Proceedings of the 4th International Symposium on Information, Computer, and Communications Security (2009),
171--182.

\end{thebibliography}
\end{document}